\documentclass[twocolumn,superscriptaddress,amsmath,amssymb,aps,prb]{revtex4-2}
\usepackage{graphicx}
\usepackage{CJKutf8}
\usepackage{dcolumn}
\usepackage{amsmath,bm}
\usepackage{epstopdf}
\usepackage[mathlines]{lineno}
\usepackage{color}
\usepackage[colorlinks=true,linkcolor=blue,citecolor=magenta,urlcolor=cyan]{hyperref}
\usepackage{orcidlink}

\begin{document}
\title{Subband structure of a cylindrical HgTe nanowire: transition from normal type to inverted type}
\author{Rui\! Li~(\begin{CJK}{UTF8}{gbsn}李睿\end{CJK})\,\orcidlink{0000-0003-1900-5998}}
\email{ruili@ysu.edu.cn}
\affiliation{Hebei Key Laboratory of Microstructural Material Physics, School of Science, Yanshan University, Qinhuangdao 066004, China}

\begin{abstract}
Based on the $6\times6$ Kane model in the spherical approximation, both the electron and the hole subband dispersions in a cylindrical HgTe nanowire are calculated using analytical method. The transcendental equations determining the subband energies in the nanowire are analytically derived by utilizing both the total angular momentum conservation and the hard wall boundary condition. We show there exists topological band transition in the HgTe nanowire, very similar to that demonstrated in a HgTe quantum well. Via tuning the radius of the nanowire, we find there is a gap-closing-and-reopening transition in the subband structure. When the radius is larger than the critical radius $R_{\rm c}\approx3.2$ nm, the HgTe nanowire is in the inverted regime. 
\end{abstract}
\date{\today}
\maketitle

\section{Introduction}

Topological insulator is a new state of matter that has a bulk band gap similar to an ordinary insulator and at the same time has conducting states on its edge or surface~\cite{RevModPhys.82.3045,RevModPhys.83.1057,bernevig2013topological,shen2012topological}. Haldane's model for realizing the quantum Hall effect in the absence of magnetic field was regarded as a model of Chern insulator~\cite{PhysRevLett.61.2015}, where the topological invariant is called the Thouless-Kohmoto-Nightingale-Nijs number or the Chern number~\cite{PhysRevLett.49.405}. There is no time reversal symmetry in Haldane's model. Later, Kane and Mele generalized Haldane's model to the case of including the spin degree of freedom~\cite{PhysRevLett.95.226801,PhysRevLett.95.146802}. The Kane-Mele model can be regarded as two spin copes of Haldane model, and is time reversal invariant. The topological invariant in this case is characterized by the $Z_{2}$ index~\cite{PhysRevLett.95.146802}. 

In 2006, Bernevig, Hughes, and Zhang proposed to look for the topological insulator phase in a HgTe/CdTe quantum well~\cite{doi:10.1126/science.1133734}. HgTe is a material with a negative band gap, while CdTe has a positive band gap~\cite{madelung2004semiconductors}. Via tuning the well width, the subband structure of the quantum well exhibits gap-closing-and-reopening transition. When the well width is larger than a critical value, the quantum well is in a topological insulator phase with a pair of helical edge states~\cite{doi:10.1126/science.1133734,bernevig2013topological,doi:10.1143/JPSJ.77.031007}. This theoretical prediction was confirmed immediately by experiment~\cite{doi:10.1126/science.1148047}. Similar band inversion transition was proposed to exist in a InAs/GaSb/AlSb quantum well~\cite{PhysRevLett.100.236601}. This topological band inversion transition is robust even in the presence of bulk inversion asymmetry~\cite{PhysRevB.77.125319,WINKLER20122096,PhysRevB.91.081302}. 

It is an interesting question whether this topological band inversion transition still survives in a HgTe nanowire. Note that quasi-one-dimensional HgTe nanowires or core/shell heterostructures are achievable using current experimental techniques ~\cite{PhysRevMaterials.1.023401,PhysRevResearch.4.023114,PhysRevMaterials.4.066001}. Although a quasi-one-dimensional nanowire is geometrically different from a quasi-two-dimensional quantum well, it is highly likely both systems share the same kind of topological band inversion transition.  After all, the negative band gap of the HgTe material seems to be the only essential factor in this topological transition~\cite{doi:10.1126/science.1133734}. Here we seek this topological band transition in a cylindrical HgTe nanowire. Also, for simplicity, we replace the barrier material CdTe with the vacuum, i.e., the HgTe nanowire is surrounded by a trivial insulator with an infinite band gap. 

Following the method introduced by Sercel and Vahala~\cite{PhysRevB.42.3690}, we rewrite the $6\times6$ Kane model~\cite{KANE1957249}, i.e., the ${\bf k}\cdot{\bf p}$ model involving both the $\Gamma_{6}$ and the $\Gamma_{8}$ bands, in the cylindrical coordinate representation. Both the bulk spectrum and bulk wave functions are obtained in this representation. The transcendental equations determining the subband dispersions are obtainable in terms of the bulk solutions. Especially, the transcendental equations at $k_{z}R=0$ are explicitly given and discussed. Clear evidence of the band inversion transition, i.e., the gap-closing-and-reopening transition, at a critical nanowire radius is given. Also, the complete subband dispersions near this critical radius are calculated. We emphasize that in our calculations, we have used the spherical approximation in the $\Gamma_{8}$ band block of the Kane model~\cite{PhysRev.102.1030,winkler2003spin}.

\section{\label{sec_kanemodel}Kane model in the cylindrical coordinate representation}
\begin{table}
\caption{\label{tab_bandpars}Bulk band parameters of HgTe material~\cite{PhysRevB.63.245305,PhysRevB.72.035321}.}
\begin{ruledtabular}
\begin{tabular}{cccccc}
$\frac{\hbar^{2}\varepsilon_{0}}{2m_{0}}$\footnote{$m_{0}$ is the free electron mass}&$\frac{\hbar^{2}P^{2}}{2m_{0}}$&$\gamma_{1}$&$\gamma_{2}$&$\gamma_{3}$&$\gamma_{s}=\frac{2\gamma_{2}+3\gamma_{3}}{5}$\\
-0.303 eV&18.8 eV&4.1&0.5&1.3&0.98
\end{tabular}
\end{ruledtabular}
\end{table}

\begin{figure}
\includegraphics{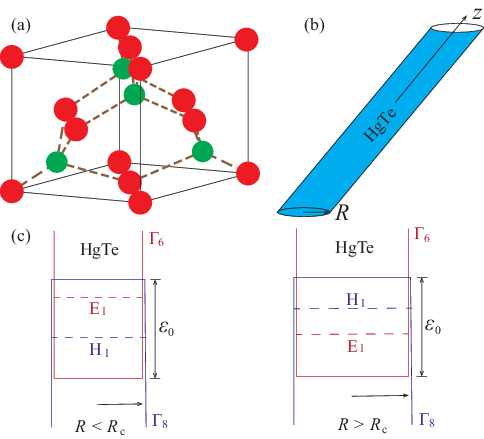}
\caption{\label{fig_nanowire}(a) Unit cell of bulk HgTe. Hg and Te atoms are marked in red and green, respectively. (b) A cylindrical HgTe nanowire with radius $R$ is under investigation. (c) Band edge profile of the nanowire. Nanowire in the normal regime $E_{1}>H_{1}$ with $R<R_{c}$ and in the reverted regime $E_{1}<H_{1}$ with $R>R_{c}$. Here $E_{1}$ and $H_{1}$ are the lowest electron subband and the highest hole subband, respectively.}
\end{figure}

{\color{red}}
Bulk HgTe has a zincblende crystal structure [see Fig.~\ref{fig_nanowire}(a)] and is known to have a negative band gap, i.e., the $\Gamma_{6}$ band of HgTe lies below its $\Gamma_{8}$ band [see Fig.~\ref{fig_nanowire}(c)]~\cite{madelung2004semiconductors}. It follows that the minimal bulk model of HgTe should be a $6\times6$ Kane model, that contains the information of the $\Gamma_{6}$ band, the $\Gamma_{8}$ band, and the coupling between them~\cite{KANE1957249}. Previously, the subband structure of a HgTe/CdTe nanowire with a rectangular cross section had been studied numerically, and the Rashba spin splitting was calculated as well~\cite{PhysRevB.96.115443}. Here we study analytically a cylindrical HgTe nanowire [see Fig.~\ref{fig_nanowire}(b)] and focus on the tunability of the subband structure via varying the nanowire radius [see Fig.~\ref{fig_nanowire}(c)]. The nanowire axis is defined as the $z$ axis, and the other two axes $x$ and $y$ are perpendicular to the nanowire. We employ the $6\times6$ Kane model~\cite{KANE1957249} in the spherical approximation in our study
\begin{widetext}
\begin{equation}
H_{0}=\frac{\hbar^{2}}{2m_{0}}\left(\begin{array}{cccccc}
\varepsilon_{0}+{\bf k}^{2}&0&-\frac{1}{\sqrt{2}}Pk_{+}&\sqrt{\frac{2}{3}}Pk_{z}&\frac{1}{\sqrt{6}}Pk_{-}&0\\
0&\varepsilon_{0}+{\bf k}^{2}&0&-\frac{1}{\sqrt{6}}Pk_{+}&\sqrt{\frac{2}{3}}Pk_{z}&\frac{1}{\sqrt{2}}Pk_{-}\\
-\frac{1}{\sqrt{2}}Pk_{-}&0&H_{0}^{33}&2\sqrt{3}\gamma_{s}k_{z}k_{-}&\sqrt{3}\gamma_{s}k^{2}_{-}&0\\
\sqrt{\frac{2}{3}}Pk_{z}&-\frac{1}{\sqrt{6}}Pk_{-}&2\sqrt{3}\gamma_{s}k_{z}k_{+}&H_{0}^{44}&0&\sqrt{3}\gamma_{s}k^{2}_{-}\\
\frac{1}{\sqrt{6}}Pk_{+}&\sqrt{\frac{2}{3}}Pk_{z}&\sqrt{3}\gamma_{s}k^{2}_{+}&0&H_{0}^{55}&-2\sqrt{3}\gamma_{s}k_{z}k_{-}\\
0&\frac{1}{\sqrt{2}}Pk_{+}&0&\sqrt{3}\gamma_{s}k^{2}_{+}&-2\sqrt{3}\gamma_{s}k_{z}k_{+}&H_{0}^{66}\end{array}\right),
\end{equation}
\end{widetext}
where $k_{\pm}=k_{x}\pm\,ik_{y}$ and
\begin{eqnarray}
H_{0}^{33}&=&H_{0}^{66}=-\big[(\gamma_{1}+\gamma_{s})(k^{2}_{x}+k^{2}_{y})+(\gamma_{1}-2\gamma_{s})k^{2}_{z}\big],\nonumber\\
H_{0}^{44}&=&H_{0}^{55}=-\big[(\gamma_{1}-\gamma_{s})(k^{2}_{x}+k^{2}_{y})+(\gamma_{1}+2\gamma_{s})k^{2}_{z}\big].\label{eq_Kaneelement}
\end{eqnarray}
Here $\varepsilon_{0}$, $P$, $\gamma_{1}$, and $\gamma_{s}$ are bulk band parameters of the HgTe material (see Tab.~\ref{tab_bandpars}~\cite{PhysRevB.63.245305,PhysRevB.72.035321}). Generally speaking, the Luttinger parameters $\gamma_{2}$ and $\gamma_{3}$ in the $\Gamma_{8}$ band are not equal. However, the spherical approximation (replacing $\gamma_{2,3}$ with $\gamma_{s}$) in most cases does not change the qualitative properties of a given system~\cite{RL2023c}. In particular, the amount of calculations including both the analytical and the numerical are greatly reduced by the spherical approximation. Considering the complexity of the size quantization of a nanowire involving the $6\times6$ bulk Hamiltonian, here we adopt the spherical approximation where the new parameter $\gamma_{s}$ is set to $\gamma_{s}=(2\gamma_{2}+3\gamma_{3})/5$~\cite{PhysRevB.8.2697,PhysRevB.84.195314}.

The advantage of taking the spherical approximation can be seen right away. The bulk Hamiltonian $H_{0}$ now commutes with the $z$-component of the total angular momentum $F_{z}$~\cite{PhysRevB.42.3690}, where
\begin{equation}
F_{z}=-i\partial_{\varphi}+\left(\begin{array}{cc}s_{z}&0_{2\times4}\\0_{4\times2}&J_{z}\end{array}\right),
\end{equation}
with $s_{z}$ and $J_{z}$ being the $z$-components of the standard spin-$1/2$ (for electron) and spin-$3/2$ (for hole) operators, respectively. Note that here we have chosen to study in a cylindrical coordinate representation where $x=r\cos\varphi$, $y=r\sin\varphi$, and $z=z$. The conservation of $F_{z}$ indicates the Hilbert space of $H_{0}$ can be divided into a series of subspace. The Hilbert subspace specified with a general value of $F_{z}=m+1/2$ ($m=0,\pm1,\pm2\ldots$) is spanned by $J_{m}(\mu\,r)e^{im\varphi}|1/2\rangle_{\rm e}$, $J_{m+1}(\mu\,r)e^{i(m+1)\varphi}|-1/2\rangle_{\rm e}$, $J_{m-1}(\mu\,r)e^{i(m-1)\varphi}|3/2\rangle_{\rm h}$, $J_{m}(\mu\,r)e^{im\varphi}|1/2\rangle_{\rm h}$, $J_{m+1}(\mu\,r)e^{i(m+1)\varphi}|-1/2\rangle_{\rm h}$, and $J_{m+2}(\mu\,r)e^{i(m+2)\varphi}|-3/2\rangle_{\rm h}$, where $J_{m}(\mu\,r)$ is the $m$-order Bessel function of the first kind~\cite{zhuxi_wang}. This Hilbert subspace has a dimension of six, such that we can rewrite $H_{0}$ as a $6\times6$ matrix in this Hilbert subspace. After some tedious derivations similar to that in Refs.~\cite{PhysRevB.42.3690,RL_2024_arxiv}, we have
\begin{widetext}
\begin{equation}
H_{0}=\frac{\hbar^{2}}{2m_{0}}\left(\begin{array}{cccccc}
\varepsilon_{0}+\mu^{2}+k^{2}_{z}&0&-\frac{i}{\sqrt{2}}P\mu&\sqrt{\frac{2}{3}}Pk_{z}&-\frac{i}{\sqrt{6}}P\mu&0\\
0&\varepsilon_{0}+\mu^{2}+k^{2}_{z}&0&-\frac{i}{\sqrt{6}}P\mu&\sqrt{\frac{2}{3}}Pk_{z}&-\frac{i}{\sqrt{2}}P\mu\\
\frac{i}{\sqrt{2}}P\mu&0&H_{0}^{33}&-2\sqrt{3}i\gamma_{s}k_{z}\mu&-\sqrt{3}\gamma_{s}\mu^{2}&0\\
\sqrt{\frac{2}{3}}Pk_{z}&\frac{i}{\sqrt{6}}P\mu&2\sqrt{3}i\gamma_{s}k_{z}\mu&H_{0}^{44}&0&-\sqrt{3}\gamma_{s}\mu^{2}\\
\frac{i}{\sqrt{6}}P\mu&\sqrt{\frac{2}{3}}Pk_{z}&-\sqrt{3}\gamma_{s}\mu^{2}&0&H_{0}^{55}&2\sqrt{3}i\gamma_{s}k_{z}\mu\\
0&\frac{i}{\sqrt{2}}P\mu&0&-\sqrt{3}\gamma_{s}\mu^{2}&-2\sqrt{3}i\gamma_{s}k_{z}\mu&H_{0}^{66}\end{array}\right),\label{eq_h0_cy}
\end{equation}
\end{widetext}
where $H^{33}_{0}$, $H^{44}_{0}$, $H^{55}_{0}$, and $H^{66}_{0}$ have the same forms as that given by Eq.~(\ref{eq_Kaneelement}) in which a simple replacement $k^{2}_{x}+k^{2}_{y}=\mu^{2}$ should be made.

\begin{figure}
\includegraphics{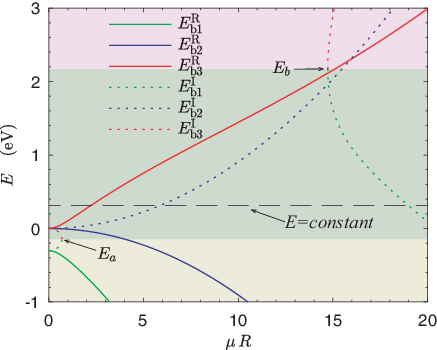}
\caption{\label{fig_bulk1}The bulk band dispersions of HgTe at a fixed longitudinal wave vector $k_{z}R=0$. The energies of the two touching points between the $E^{\rm I}_{\rm b1}$ and $E^{\rm I}_{\rm b3}$ branches are given by $E_{a}$ and $E_{b}$, respectively. Here the nanowire radius is chosen as $R=3$ nm. We divide the whole energy region into three subregions $E<E_{\rm a}$, $E_{\rm a}<E<E_{\rm b}$, and $E_{\rm b}<E$.}
\end{figure}

\begin{figure}
\includegraphics{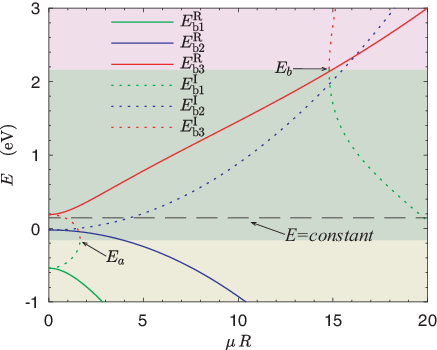}
\caption{\label{fig_bulk2}The bulk band dispersions of HgTe at a fixed longitudinal wave vector $k_{z}R=1.5$. The two energies $E_{a}$ and $E_{b}$ have the same meaning as that in Fig.~\ref{fig_bulk1}, and the nanowire radius is $R=3$ nm. We divide the whole energy region into three subregions $E<E_{\rm a}$, $E_{\rm a}<E<E_{\rm b}$, and $E_{\rm b}<E$.}
\end{figure}

Diagonalizing the bulk Hamiltonian (\ref{eq_h0_cy}) in the cylindrical coordinate representation, we obtain three branches of bulk dispersion 
\begin{eqnarray}
E^{\rm R}_{\rm b1}&=&\frac{\varepsilon_{0}}{2}-\frac{1}{2}(\gamma_{1}+2\gamma_{s}-1)(\mu^{2}+k^{2}_{z})-\frac{1}{6}\chi_{\mu},\nonumber\\
E^{\rm R}_{\rm b2}&=&-(\gamma_{1}-2\gamma_{s})(\mu^{2}+k^{2}_{z}),\nonumber\\
E^{\rm R}_{\rm b3}&=&\frac{\varepsilon_{0}}{2}-\frac{1}{2}(\gamma_{1}+2\gamma_{s}-1)(\mu^{2}+k^{2}_{z})+\frac{1}{6}\chi_{\mu},\label{eq_spectrumbulk}
\end{eqnarray}
where
\begin{eqnarray}
\chi_{\mu}&=&\Big[12(\mu^{2}+k^{2}_{z})\Big(3(\varepsilon_{0}+\mu^{2}+k^{2}_{z})(\gamma_{1}+2\gamma_{s})+2P^{2}\Big)\nonumber\\
&&+9\Big(\varepsilon_{0}-(\gamma_{1}+2\gamma_{s}-1)(\mu^{2}+k^{2}_{z})\Big)^{2}\Big]^{1/2}.
\end{eqnarray}
Each branch is two-fold degenerate, i.e., there exist two bulk wave functions for each bulk dispersion (for details see appendix~\ref{appendix_bulkwavefun}). Note that we have implicitly assumed $\mu$ to be real in the bulk dispersions given by Eq.~(\ref{eq_spectrumbulk}). However, in order to the solve the suband quantization under a transverse hard wall confining potential, we also need to consider Bessel functions with imaginary argument $i\mu\,r$~\cite{zhuxi_wang}. Replacing $\mu$ with $i\mu$ in Eq.~(\ref{eq_spectrumbulk}), we have the bulk dispersions in terms of the Bessel functions with imaginary argument~\cite{Li-lirui:2024we} 
\begin{equation}
E^{\rm I}_{\rm b1,2,3}=\left.E^{\rm R}_{\rm b1,2,3}\right|_{\mu=i\mu}.
\end{equation}
 
We now have the complete bulk dispersions governed by the bulk Hamiltonian (\ref{eq_h0_cy}). The complete bulk dispersions with the longitudinal wave vector fixed at $k_{z}R=0$ and $k_{z}R=1.5$ are shown in Figs.~\ref{fig_bulk1} and \ref{fig_bulk2}, respectively. One can find that a given constant energy line always intersects the bulk dispersions three times. This means we always can write the eigenfunction in the nanowire as a linear combination of six bulk wave functions. Each intersecting point gives rise to two bulk wave functions, and three intersecting points correspond to six bulk wave functions. Also, from the bulk dispersions shown in Figs.~\ref{fig_bulk1} and \ref{fig_bulk2}, we find the whole energy region $-\infty<E<\infty$ can always be divided into three subregions, i.e., $E<E_{\rm a}$, $E_{\rm a}<E<E_{\rm b}$, and $E_{\rm b}<E$. The situation of the intersecting between the constant energy line and the bulk dispersions varies with the subregion. Therefore, we need to select the proper bulk wave functions in each energy subregion.

\section{\label{sec_subband_edge}Subband energies at $k_{z}R=0$}
{\color{red}}
Topological band inversion transition has been previously shown to occur in a HgTe quantum well where the CdTe material serves as the confining potential~\cite{doi:10.1126/science.1133734}. Here for simplicity, we consider a bare cylindrical HgTe nanowire, i.e., the vacuum serves as the confining potential. The vacuum can be considered as a trivial insulator with an infinite band gap. The Hamiltonian governing the subband quantization reads
\begin{equation}
H=H_{0}+V(r),\label{eq_subbandquan}
\end{equation}
where 
\begin{equation}
V(r)=\left\{\begin{array}{cc}0_{6\times6},& r<R,\\\infty\left(\begin{array}{cc}1_{2\times2}&0_{2\times4}\\0_{4\times2}&-1_{4\times4}\end{array}\right),&r>R,\end{array}\right.
\end{equation}
with $R$ being the radius of the nanowire. Here, we have a $+\infty$ confining potential for electron in the $\Gamma_{6}$ band and a $-\infty$ confining potential for hole in the $\Gamma_{8}$ band. In other words, we have hard wall boundary condition both in the $\Gamma_{6}$ and $\Gamma_{8}$ bands [see Fig.~\ref{fig_nanowire}(c)]. Because of the zincblende structure of HgTe [see Fig.~\ref{fig_nanowire}(a)], nanowire grown along the [001]/[111] direction should have a square/hexagonal cross section. Anyway, it is easier to study a nanowire with a circular cross section and its results can reflect the qualitative properties of various realistic nanowires.


There is no confinement in the $z$ direction, such that $k_{z}$ is a conserved quantity in Hamiltonian (\ref{eq_subbandquan}). It follows that the eigenvalues of Hamiltonian (\ref{eq_subbandquan}) can always be written as a quasi-one-dimensional dispersion $E(k_{z})$. Here we first focus on the subband energies at $k_{z}R=0$. The bulk Hamiltonian (\ref{eq_h0_cy}) is block diagonalized at $k_{z}R=0$, i.e., we can regroup the bulk Hamiltonian to a form with two nonvanishing $3\times3$ blocks. It is likely to have relatively simple results at this site. The block diagonalized feature of $H_{0}$ indicates we have two independent Hilbert subspaces, i.e., one subspace I spanned by $J_{m}(\mu\,r)e^{im\varphi}|1/2\rangle_{\rm e}$, $J_{m-1}(\mu\,r)e^{i(m-1)\varphi}|3/2\rangle_{\rm h}$, and $J_{m+1}(\mu\,r)e^{i(m+1)\varphi}|-1/2\rangle_{\rm h}$, and the other subspace II spanned by  $J_{m+1}(\mu\,r)e^{i(m+1)\varphi}|-1/2\rangle_{\rm e}$, $J_{m}(\mu\,r)e^{im\varphi}|1/2\rangle_{\rm h}$, and $J_{m+2}(\mu\,r)e^{i(m+2)\varphi}|-3/2\rangle_{\rm h}$. Note that similar classification of the Hilbert subspaces has been used in studying the interface state between inverted and normal semiconductors~\cite{PhysRevB.105.035305}. In the following, we separately look for the subband energies in these two Hilbert subspaces. 

Writing the eigenfunction as a linear combination of three bulk wave functions in each of the Hilbert subspaces, and imposing the hard wall boundary condition at $r=R$,  we obtain one transcendental equation in the Hilbert subspace I 
\begin{equation}
\left|\begin{array}{ccc}
M^{\pm}_{11}&M^{\pm}_{12}&0\\
\sqrt{3}J_{m-1}(\mu_{1}R)&\sqrt{3}J_{m-1}(\mu_{3}R)&-\frac{1}{\sqrt{3}}J_{m-1}(\mu_{2}R)\\
J_{m+1}(\mu_{1}R)&J_{m+1}(\mu_{3}R)&J_{m+1}(\mu_{2}R)
\end{array}\right|=0,\label{eq_transc_edge}
\end{equation}
and the other transcendental equation in the Hilbert subspace II
\begin{equation}
\left|\begin{array}{ccc}
O^{\pm}_{11}&O^{\pm}_{12}&0\\
\frac{1}{\sqrt{3}}J_{m}(\mu_{1}R)&\frac{1}{\sqrt{3}}J_{m}(\mu_{3}R)&-\sqrt{3}J_{m}(\mu_{2}R)\\
J_{m+2}(\mu_{1}R)&J_{m+2}(\mu_{3}R)&J_{m+2}(\mu_{2}R)
\end{array}\right|=0,\label{eq_transc_edge2}
\end{equation}
where 
\begin{eqnarray}
M^{\pm}_{11}&=&-i\frac{3\varepsilon_{0}+3(\gamma_{1}+2\gamma_{s}+1)\mu^{2}_{1}\pm\chi_{\mu_{1}}}{\sqrt{6}P\mu_{1}}J_{m}(\mu_{1}R),\nonumber\\
M^{\pm}_{12}&=&-i\frac{3\varepsilon_{0}+3(\gamma_{1}+2\gamma_{s}+1)\mu^{2}_{3}\pm\chi_{\mu_{3}}}{\sqrt{6}P\mu_{3}}J_{m}(\mu_{3}R),\nonumber\\
O^{\pm}_{11}&=&-i\frac{3\varepsilon_{0}+3(\gamma_{1}+2\gamma_{s}+1)\mu^{2}_{1}\pm\chi_{\mu_{1}}}{3\sqrt{2}P\mu_{1}}J_{m+1}(\mu_{1}R),\nonumber\\
O^{\pm}_{12}&=&-i\frac{3\varepsilon_{0}+3(\gamma_{1}+2\gamma_{s}+1)\mu^{2}_{3}\pm\chi_{\mu_{3}}}{3\sqrt{2}P\mu_{3}}J_{m+1}(\mu_{3}R),
\end{eqnarray}
and
\begin{widetext}
{\small
\begin{eqnarray}
\mu_{1/3}&=&\frac{\left(\mp\sqrt{\big(3(\varepsilon_{0}-E)(\gamma_{1}+2\gamma_{s})+3E+2P^2\big)^2-36E(\varepsilon_{0}-E)(\gamma_{1}+2\gamma_{s})}-3\varepsilon_{0}(\gamma_{1}+2\gamma_{s})+3E(\gamma_{1}+2\gamma_{s}-1)-2 P^2\right)^{1/2}}{\sqrt{6(\gamma_{1}+2\gamma_{s})}},\nonumber\\
\mu_{2}&=&\sqrt{\frac{-E}{(\gamma_{1}-2\gamma_{s})}}.
\end{eqnarray}
}
\end{widetext}
Here $\mu_{1,3}$ are solved from $E_{\rm b1,3}$ and $\mu_{2}$ is solved from $E_{\rm b2}$ [see Eq.~(\ref{eq_spectrumbulk})]. Note that the subband energy $E$ is the only unknown in Eqs.~(\ref{eq_transc_edge}) and (\ref{eq_transc_edge2}), and how to select the proper matrix elements $M^{\pm}_{11}$, $M^{\pm}_{12}$, $O^{\pm}_{11}$, and $O^{\pm}_{12}$ will be explained in the following.

\begin{figure}
\includegraphics{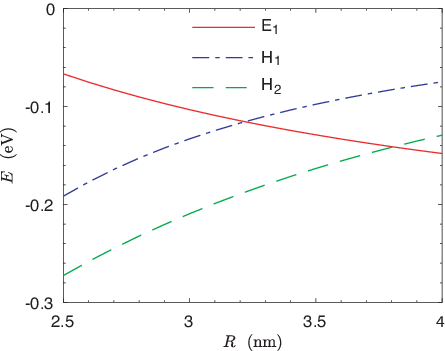}
\caption{\label{fig_bandedgeenergy}The subband energies at $k_{z}R=0$ as a function of the nanowire radius $R$. The lowest electron subband insects with the two top most hole subbands both at $R\approx3.2$ nm and $R\approx3.8$ nm. The fundamental band gap closes at the critical radius $R_{\rm c}\approx3.2$ nm.}
\end{figure}

Let us discuss the special case $|F_{z}|=1/2$. This case is important because the subband energies at the fundamental band gap are given by the smallest total angular momentum $|F_{z}|=1/2$~\cite{PhysRevB.42.3690,sweeny1988hole,PhysRevB.79.155323,PhysRevB.84.195314,Li-lirui:2024we,RL_2024_arxiv}, i.e., $m=0$ and $m=-1$. For the case $m=0$, using the property of the Bessel function $J_{-1}(\mu_{2}R)=-J_{1}(\mu_{2}R)$~\cite{zhuxi_wang}, the transcendental equation (\ref{eq_transc_edge}) can be further reduced to two independent transcendental equations
\begin{equation}
\left|\begin{array}{cc}
\left.M^{\pm}_{11}\right|_{m=0}&\left.M^{\pm}_{12}\right|_{m=0}\\
J_{1}(\mu_{1}R)&J_{1}(\mu_{3}R)
\end{array}\right|=0,~~{\rm and}~~J_{1}(\mu_{2}R)=0.\label{eq_transc_edge3}
\end{equation}
For the case $m=-1$, the transcendental equation (\ref{eq_transc_edge2}) can be further reduced to two independent transcendental equations
\begin{equation}
\left|\begin{array}{cc}
\left.O^{\pm}_{11}\right|_{m=-1}&\left.O^{\pm}_{12}\right|_{m=-1}\\
J_{1}(\mu_{1}R)&J_{1}(\mu_{3}R)
\end{array}\right|=0,~~{\rm and}~~J_{1}(\mu_{2}R)=0.\label{eq_transc_edge4}
\end{equation}
The above two equations (\ref{eq_transc_edge3}) and (\ref{eq_transc_edge4}) are very important for understanding the band inversion transition at a critical radius.  

We now explain how to select the proper matrix elements in the transcendental equations (\ref{eq_transc_edge}) and (\ref{eq_transc_edge2}). This selection is intimately relevant to the region division shown in Fig.~\ref{fig_bulk1}. In energy subregion $E<E_{a}$, we should select the matrix elements $M^{-}_{11}$ and $M^{-}_{12}$ in Eq.~(\ref{eq_transc_edge}), and $O^{-}_{11}$ and $O^{-}_{12}$ in Eq.~(\ref{eq_transc_edge2}). In energy subregion $E_{a}<E<E_{b}$, we should select the matrix elements $M^{-}_{11}$ and $M^{+}_{12}$ in Eq.~(\ref{eq_transc_edge}), and $O^{-}_{11}$ and $O^{+}_{12}$ in Eq.~(\ref{eq_transc_edge2}). In energy subregion $E_{b}<E$, we should select the matrix elements $M^{+}_{11}$ and $M^{+}_{12}$ in Eq.~(\ref{eq_transc_edge}), and $O^{+}_{11}$ and $O^{+}_{12}$ in Eq.~(\ref{eq_transc_edge2}).

Although we have derived the transcendental equations of the subband energy $E$ at $k_{z}R=0$ in the whole energy region $-\infty<E<\infty$, the most interesting subband energies actually lie in the band gap of HgTe~\cite{doi:10.1126/science.1133734}, i.e., $-0.303~{\rm eV}<E<0$ eV. Also, both the lowest electron subband and the top most hole subbands are given by the $|F_{z}|=1/2$, i.e., $m=0$ and $m=-1$. If we are only interested in the subband energies near the fundamental band gap, we just need to solve Eqs.~(\ref{eq_transc_edge3}) and (\ref{eq_transc_edge4}). The lowest electron subband and the two top most hole subbands as a function of the nanowire radius are shown in Fig.~\ref{fig_bandedgeenergy}. We find the lowest electron subband intersects with the two top most hole subbands both at $R\approx3.2$ nm and $R\approx3.8$ nm. The radius $R\approx3.2$ nm can be regarded as the critical radius $R_{\rm c}$, at which the fundamental band gap closes. 

The negative bulk band gap $\varepsilon_{0}$ of HgTe plays an essential role in the gap-closing-and-reopening transition [see Fig.~\ref{fig_nanowire}(c)]. The bulk band gap $\varepsilon_{0}$ can be tunable to some extent, e.g., via changing the composition $x$ in the compound ${\rm Hg}_{1-x}{\rm Cd}_{x}{\rm Te}$~\cite{10.1063/1.345119,Orlita:2014tw}. Here, we discuss the potential dependence of the critical radius $R_{\rm c}$ on the bulk band gap $\varepsilon_{0}$. The subband energy $E_{1}/H_{1}$ decreases/increases with the radius $R$ (see Fig.~\ref{fig_bandedgeenergy}), such that more negative $\varepsilon_{0}$ will lead to smaller critical radius $R_{\rm c}$ [see Fig.~\ref{fig_nanowire}(c)]. In other words, with the increase of $x$ in the compound ${\rm Hg}_{1-x}{\rm Cd}_{x}{\rm Te}$, $R_{\rm c}$ also increases and eventually becomes infinite at $x=0.17$ where the bulk band gap vanishes~\cite{Orlita:2014tw}. Very similar band inversion transition was reported in a SnTe nanowire with a  square cross section~\cite{doi:10.1021/acsanm.4c00506}. The critical nanowire size in that case was estimated to be 17 nm~\cite{doi:10.1021/acsanm.4c00506}.  Because SnTe has a less negative band gap $\varepsilon_{0}$ than HgTe, it is reasonable HgTe nanowire has a smaller critical size $2R_{\rm c}\approx6.4$ nm here. Also, the geometry of the nanowire cross section may have influence on the critical size. We emphasize the critical size in the HgTe nanowire is almost identical to that in a HgTe quantum well where $d_c=64$ \AA~\cite{doi:10.1126/science.1133734}.

\section{Subband dispersions}
{\color{red}}
We now continue to study the subband energies at a general wave vector site $k_{z}R\neq0$, i.e., we want to obtain the subband energies as a function of the longitudinal wave vector $k_{z}$. Similar to the case of $k_{z}R=0$, here we still can derive a series of transcendental equations with respect to the region division. When $k_{z}R\neq0$, from the bulk spectrum shown in Fig.~\ref{fig_bulk2}, we find the whole energy region can still be divided into three subregions. 

In energy subregion $E<E_{a}$, a constant energy line intersects the bulk dispersion $E_{\rm b1}$ two times and the bulk dispersion $E_{\rm b2}$ one time, such that the eigenfunction is expanded in terms of four bulk wave functions from the $E_{\rm b1}$ branch and two bulk wave functions from the $E_{\rm b2}$ branch. In energy subregion $E_{a}<E<E_{b}$, a constant energy line intersects each of the bulk dispersions $E_{\rm b1,2,3}$ just one time, such that the eigenfunction is expanded in terms of two bulk wave functions from the $E_{\rm b1}$ branch, two bulk wave functions from the $E_{\rm b2}$ branch, and two bulk wave functions from the $E_{\rm b3}$ branch. In energy subregion $E_{b}<E$, a constant energy line intersects the bulk dispersion $E_{\rm b3}$ two times and the bulk dispersion $E_{\rm b2}$ one time, such that the eigenfunction is expanded in terms of four bulk wave functions from the $E_{\rm b3}$ branch and two bulk wave functions from the $E_{\rm b2}$ branch. Letting the eigenfunction vanish at the boundary $\Psi(R,\varphi,z)=0$, we are able to obtain six equations of the expansion coefficients. Letting the determinant of the coefficient matrix equal to zero, we have the transcendental equation in each of the three energy subregions. 

Due to the tedious expression of the bulk wave functions (see appendix~\ref{appendix_bulkwavefun}), here we do not show these $6\times6$ coefficient matrices in three energy subregions. The procedure of deriving these $6\times6$ coefficient matrices is exact the same as that in the hole subband calculations of a Ge nanowire~\cite{,Li-lirui:2024we,RL_2024_arxiv}. Solving these three transcendental equations, we are able to have the complete subband dispersions $E(k_{z})$. Certainly, our method is self-consistent. At the site $k_{z}R=0$, these transcendental equations, i.e., equalling the determinant of the $6\times6$ coefficient matrices to zero, can be reduced to two independent transcendental equations given by Eqs.~(\ref{eq_transc_edge}) and (\ref{eq_transc_edge2}).

\begin{figure}
\includegraphics{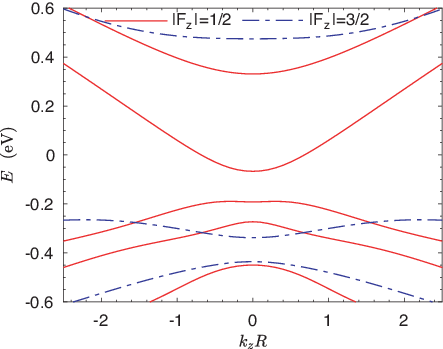}
\caption{\label{fig_Subband2point5}Subband dispersions in a cylindrical HgTe nanowire with radius $R=2.5$ nm. The electron subband $E_{1}$ lies above the hole subband $H_{1}$. The nanowire is in the normal regime. }
\end{figure}

\begin{figure}
\includegraphics{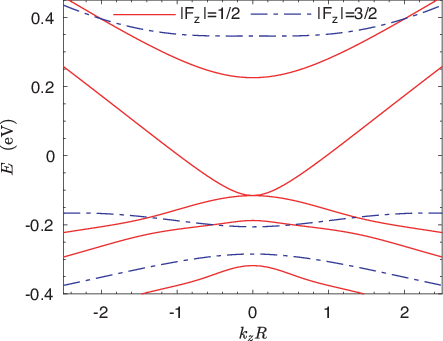}
\caption{\label{fig_Subband3point2}Subband dispersions in a cylindrical HgTe nanowire with radius $R=3.2$ nm. The electron subband $E_{1}$ touches with the hole subband $H_{1}$ at $k_{z}R=0$, the band gap vanishes. The nanowire is in the critical regime.}
\end{figure}

\begin{figure}
\includegraphics{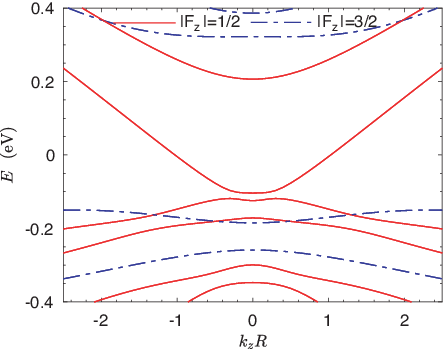}
\caption{\label{fig_Subband3point4}Subband dispersions in a cylindrical HgTe nanowire with radius $R=3.4$ nm. The electron subband $E_{1}$ lies below the hole subband $H_{1}$, and the anticrossings between subbands $E_{1}$ and $H_{1}$ give rise to a band gap.  The nanowire is in the inverted regime.}
\end{figure}

Since we have shown evidence of the band inversion transition at a critical radius via calculating the subband energies at $k_{z}R=0$ in Sec.~\ref{sec_subband_edge}, here we show the complete subband dispersions near this transition point. The subband dispersions in a cylindrical HgTe nanowire with radii $R=2.5$ nm, $R=3.2$ nm, and $R=3.4$ nm are shown in Figs.~\ref{fig_Subband2point5}, \ref{fig_Subband3point2}, and \ref{fig_Subband3point4}, respectively. At the critical radius $R_{\rm c}\approx3.2$ nm, the HgTe nanowire has a vanishing band gap (see Fig.~\ref{fig_Subband3point2}). For radius $R<R_{\rm c}$,  the HgTe nanowire has a normal subband structure, i.e., the lowest $E_{1}$ subband lies above the top most $H_{1}$ subband (see Fig.~\ref{fig_Subband2point5}). While for radius $R_{\rm c}<R$, the HgTe nanowire has an inverted subband structure, i.e., the lowest $E_{1}$ subband lies below the top most $H_{1}$ subband (see Fig.~\ref{fig_Subband3point4}). Now, the anticrossings between subbands $E_{1}$ and $H_{1}$ give rise to a band gap at $k_{z}R\neq0$. Both the low-energy effective Hamiltonian of a HgTe nanowire with $R>R_{\rm c}$ and whether there are edge states in this case will be studied in a separate paper. 

Although the band inversion transition is induced by dimensional tuning both in the HgTe nanowire and quantum well structures, here we discuss their differences. From the subband dispersions shown in Figs.~\ref{fig_Subband2point5}, \ref{fig_Subband3point2}, and \ref{fig_Subband3point4}, we find the effective Hamiltonian near the band gap does not take the Dirac Hamiltonian form. We do not have linear dispersions near the bad gap, while for HgTe quantum well the dispersions are linear near the bad gap~\cite{doi:10.1126/science.1133734}. This is the main difference between the nanowire and the quantum well structures.  Also, the edge state of a topological quasi-one-dimensional system is different from that of a topological quasi-two-dimensional system~\cite{bernevig2013topological,shen2012topological}. The potential applications of quasi-one-dimensional systems also differ from that of quasi-two-dimensional systems.

\section{Summary}
In summary, we have analytically solved the effective mass model of a cylindrical HgTe nanowire involving both the $\Gamma_{6}$ and $\Gamma_{8}$ bands. Both the spherical approximation and the hard-wall boundary condition are used in our calculations. A series of transcendental equations determining the subband energies are analytically derived. Both the subband energies at $k_{z}R=0$ and the complete subband dispersions are calculated and discussed. Topological band inversion transition is shown to occur at a critical radius. The HgTe nanowire has an inverted subband structure when its radius is larger than this critical radius.

\appendix
\section{\label{appendix_bulkwavefun}Bulk wave functions}
Diagonalizing the $6\times6$ bulk Hamiltonian (\ref{eq_h0_cy}) given in a cylindrical coordinate representation, we obtain six bulk wave functions. There are two bulk wave functions for each bulk dispersion shown in Eq.~(\ref{eq_spectrumbulk}). The two bulk wave functions corresponding to the bulk dispersion $E_{\rm b1}$ are given by
\begin{equation}
\left(\begin{array}{c}\frac{\sqrt{2}k_{z}\big(3\varepsilon_{0}+3(\gamma_{1}+2\gamma_{s}+1)(\mu^{2}+k^{2}_{z})-\chi_{\mu}\big)}{3P\mu^{2}}J_{m}(\mu\,r)e^{im\varphi}\\
-i\frac{3\varepsilon_{0}+3(\gamma_{1}+2\gamma_{s}+1)(\mu^{2}+k^{2}_{z})-\chi_{\mu}}{3\sqrt{2}P\mu}J_{m+1}(\mu\,r)e^{i(m+1)\varphi}\\
\frac{2ik_{z}}{\mu}J_{m-1}(\mu\,r)e^{i(m-1)\varphi}\\
\frac{4k^{2}_{z}+\mu^{2}}{\sqrt{3}\mu^{2}}J_{m}(\mu\,r)e^{im\varphi}\\
0\\
J_{m+2}(\mu\,r)e^{i(m+2)\varphi}
\end{array}\right),
\end{equation}
and
\begin{equation}
\left(\begin{array}{c}
-i\frac{3\varepsilon_{0}+3(\gamma_{1}+2\gamma_{s}+1)(\mu^{2}+k^{2}_{z})-\chi_{\mu}}{\sqrt{6}P\mu}J_{m}(\mu\,r)e^{im\varphi}\\
0\\
\sqrt{3}J_{m-1}(\mu\,r)e^{i(m-1)\varphi}\\
-\frac{2ik_{z}}{\mu}J_{m}(\mu\,r)e^{im\varphi}\\
J_{m+1}(\mu\,r)e^{i(m+1)\varphi}\\
0
\end{array}\right).
\end{equation}
The two bulk wave functions corresponding to the bulk dispersion $E_{\rm b2}$ are given by
\begin{equation}
\left(\begin{array}{c}0\\0\\\frac{2ik_{z}}{\mu}J_{m-1}(\mu\,r)e^{i(m-1)\varphi}\\-\sqrt{3}J_{m}(\mu\,r)e^{im\varphi}\\0\\J_{m+2}(\mu\,r)e^{i(m+2)\varphi}\end{array}\right),
\end{equation}
and 
\begin{equation}
\left(\begin{array}{c}0\\0\\-\frac{4k^{2}_{z}+\mu^{2}}{\sqrt{3}\mu^{2}}J_{m-1}(\mu\,r)e^{i(m-1)\varphi}\\-\frac{2ik_{z}}{\mu}J_{m}(\mu\,r)e^{im\varphi}\\J_{m+1}(\mu\,r)e^{i(m+1)\varphi}\\0\end{array}\right).
\end{equation}
The two bulk wave functions corresponding to the bulk dispersion $E_{\rm b3}$ are given by
\begin{equation}
\left(\begin{array}{c}\frac{\sqrt{2}k_{z}\big(3\varepsilon_{0}+3(\gamma_{1}+2\gamma_{s}+1)(\mu^{2}+k^{2}_{z})+\chi_{\mu}\big)}{3P\mu^{2}}J_{m}(\mu\,r)e^{im\varphi}\\
-i\frac{3\varepsilon_{0}+3(\gamma_{1}+2\gamma_{s}+1)(\mu^{2}+k^{2}_{z})+\chi_{\mu}}{3\sqrt{2}P\mu}J_{m+1}(\mu\,r)e^{i(m+1)\varphi}\\
\frac{2ik_{z}}{\mu}J_{m-1}(\mu\,r)e^{i(m-1)\varphi}\\
\frac{4k^{2}_{z}+\mu^{2}}{\sqrt{3}\mu^{2}}J_{m}(\mu\,r)e^{im\varphi}\\
0\\
J_{m+2}(\mu\,r)e^{i(m+2)\varphi}
\end{array}\right),
\end{equation}
and
\begin{equation}
\left(\begin{array}{c}
-i\frac{3\varepsilon_{0}+3(\gamma_{1}+2\gamma_{s}+1)(\mu^{2}+k^{2}_{z})+\chi_{\mu}}{\sqrt{6}P\mu}J_{m}(\mu\,r)e^{im\varphi}\\
0\\
\sqrt{3}J_{m-1}(\mu\,r)e^{i(m-1)\varphi}\\
-\frac{2ik_{z}}{\mu}J_{m}(\mu\,r)e^{im\varphi}\\
J_{m+1}(\mu\,r)e^{i(m+1)\varphi}\\
0
\end{array}\right).
\end{equation}
The eigenfunction of Hamiltonian (\ref{eq_subbandquan}) can be written as a linear combination of the above bulk wave functions. 
\bibliography{Ref_RL}
\end{document}